\documentclass[a4paper,11pt]{article}
\usepackage[utf8]{inputenc}
\usepackage[english]{babel}
\usepackage{braket}
\usepackage{xspace}
\usepackage{bbm}
\usepackage{mathdots}
\usepackage{stackrel}
\usepackage[dvipsnames]{xcolor}
\usepackage{appendix}
\usepackage{hyperref}
\hypersetup{
 colorlinks=true,
 linkcolor=blue,
 anchorcolor = blue,
 citecolor = blue,
 filecolor = blue,
 urlcolor = blue
}
\usepackage{mathtools}
\usepackage{bbm}
\usepackage{comment}
\usepackage{subfigure}
\usepackage[T1]{fontenc}               
\usepackage[left=3cm,
            right=2.5cm,
            top=2.5cm,
            bottom=3cm
            ]{geometry}                
\usepackage{graphicx}
\usepackage{mathrsfs}
\usepackage{appendix}
\usepackage{fancyhdr}
\usepackage{amsmath,                   
            amssymb,                   
            amsthm}                    

\usepackage{setspace}
\usepackage{cite}
\usepackage{cancel}
\usepackage{bm}
\usepackage[margin=30pt, bf, font=small, center, justification=justified]{caption}[2004/07/16]

\usepackage{setspace} 

\usepackage{tikz}

\title{\bf Dynamics of entanglement fluctuations and quantum Mpemba effect in the $\nu=1$ QSSEP model}

\author{Angelo Russotto$^1$, Filiberto Ares$^1$, Pasquale Calabrese$^{1}$, Vincenzo Alba$^2$}

\date{}

\begin{document} 

\maketitle
{\small
\vspace{-5mm}  \ \\
{$^{1}$}  SISSA and INFN Sezione di Trieste, via Bonomea 265, 34136 Trieste, Italy\\[-0.1cm]
\medskip
{$^{2}$}   Dipartimento di Fisica and INFN Sezione di Pisa, Largo Bruno Pontecorvo 3, 56126  Pisa, Italy\\[-0.1cm]
\medskip
}

\begin{abstract}
We study the out-of-equilibrium dynamics of entanglement fluctuations in the $\nu=1$ Quantum Symmetric Simple Exclusion Process, a free-fermion chain with hopping amplitudes that are stochastic in time but homogeneous in space. Previous work showed that the average entanglement growth after a quantum quench can be explained in terms of pairs of entangled quasiparticles performing random walks, leading to diffusive entanglement spreading. By incorporating the noise-induced statistical correlations between the quasiparticles, we extend this description to the full-time probability distribution of the entanglement entropy. Our generalized quasiparticle picture allows us to compute the average time evolution of a generic function of the reduced density matrix of a subsystem. We also apply our result to the entanglement asymmetry. This allows us to investigate the restoration of particle-number symmetry  in the dynamics from initial states with no well-defined particle number. Regarding the possible existence of the quantum Mpemba effect, our analysis indicates that its occurrence is an extremely fine-tuned phenomenon, requiring very specific conditions and therefore being rather difficult to observe in practice.
\end{abstract}

\tableofcontents

\section{Introduction}

Quantum models with a unitary evolution driven by stochastic processes offer a promising route to understanding the complex dynamics of out-of-equilibrium quantum systems. These models are particularly useful for describing many-body systems interacting with their environment, providing valuable insights into the fluctuations of quantum diffusive dynamics. While fluctuations in classical diffusive systems are well captured by Macroscopic Fluctuation Theory~\cite{mft}, the existence of an analogous universal framework at the quantum level remains unclear~\cite{bernard-21}. At the same time, random unitary dynamics are also useful for extracting and analyzing the typical, universal properties of isolated many-body systems. By introducing randomness, fine-grained, specific features of the system are filtered out, leaving only the more generic ones --- in some cases, randomness also gives analytical access to notoriously difficult problems, such as entanglement growth. Randomness can be implemented in many different ways: e.g. through random unitary circuits~\cite{nrvh-17, nvh-18, cdlc-18, jhn-18, krps-18, gh-19, z-20, mlgkp-25}, or as continuous-time stochastic Hamiltonian dynamics~\cite{znidaric-10, lashkari-13, bbj-17, onorati-17, cglp-17, knap-18, rl-18, bd-17, clbdl-23, sbn-25, crdl-25}.

An ideal candidate of the second type for quantitatively studying fluctuations in diffusive quantum dynamics is the Quantum Symmetric Simple Exclusion Process (QSSEP), introduced in Refs.~\cite{bbj-17, bbj2-19}. This is a system of free fermions hopping on a ring with hopping amplitudes between neighboring sites that are random both in space and time. Since the Hamiltonian generator of the dynamics is quadratic, the model is free-fermionic for each noise realization. This opens the possibility of employing analytic methods and performing efficient numerical simulations. Remarkably, the average dynamics of the 
state reproduces that of the SSEP, its classical counterpart~\cite{mallick-15}. Still, understanding the genuine quantum aspects of the model --- such as the fluctuations of the state dynamics or the entanglement --- is also crucial~\cite{bb-25, bbj3-20, hb-23}. The full statistics of the fluctuations of both the stationary state and the stationary entanglement entropy have been determined~\cite{bbj2-19, bp-21}. However, the full-time evolution of the entanglement is still missing. 

The exact full-time dynamics of the average entanglement entropy has been derived analytically in a generalization of the QSSEP model, recently introduced in Ref.~\cite{alba-25}, in which the hopping amplitudes are random in time but homogeneous in space; thereby, the model is translationally invariant, in contrast to the original QSSEP (see Refs.~\cite{bj-19, bj-21, jkb-20, behm-22, bjsw-25} for other extensions). One can relax this property and only partially restore spatial invariance by considering the same random hoppings every $\nu$ sites — hence the name $\nu$-QSSEP given in Ref.~\cite{alba-25}. The case $\nu=1$ corresponds to homogeneous random hoppings; for $\nu=2$, the noise in odd and even sites is different; and in the limit $\nu \to \infty$, the original QSSEP model is recovered. The dynamics of the averaged entanglement entropy in the $\nu=1$ case can be interpreted in terms of a stochastic generalization of the quasiparticle picture that explains entanglement spreading in the non-random counterpart of this model~\cite{cc-05, fc-08, ac-17, ac-18}. While in that case the quasiparticles propagate ballistically, leading to a linear growth of entanglement, here they follow a random walk in each realization, implying that entanglement spreads diffusively. Although an exact description of the average entanglement dynamics for generic $\nu$ is still lacking, the same qualitative behavior is observed.

\begin{figure*}[t]
\centering
\includegraphics[width=0.9\textwidth]{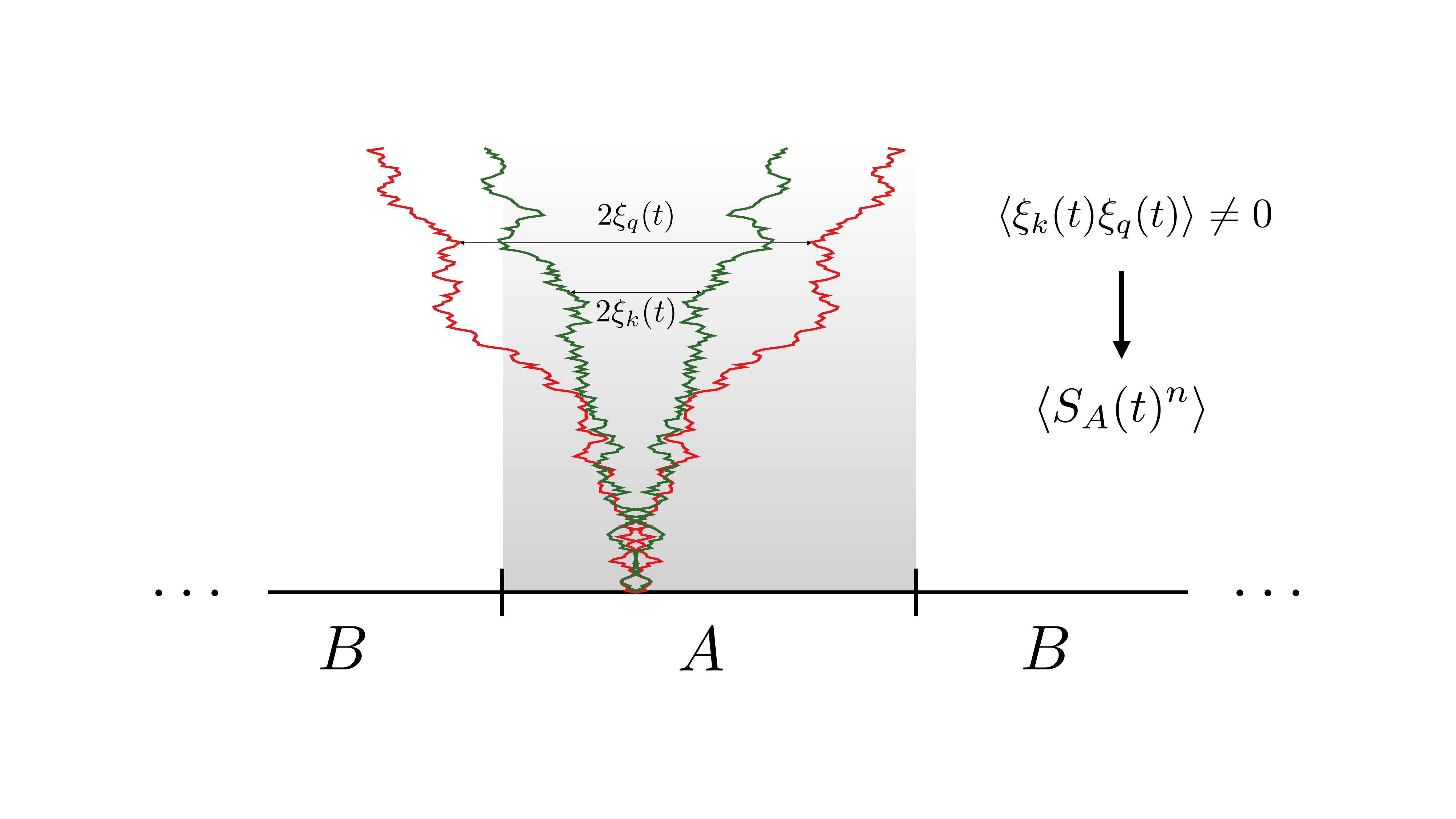}
    \caption{\label{fig:DQPsketch}Sketch of the diffusive quasiparticle picture for the out-of-equilibrium distribution of the entanglement entropy of a subsystem $A$ in the $\nu = 1$ QSSEP model~\eqref{model_dH}. At $t=0$, entangled pairs of quasiparticles, each one associated with a momentum $k$, are emitted and  propagate diffusively. The distance $\xi_k(t)$ between the two quasiparticles of a pair describes a real Brownian motion. Since $\langle \xi_k(t)\xi_q(t)\rangle\neq 0$, 
    quasiparticles with different momenta are statistically correlated. These correlations are key to describe the fluctuations of the entanglement via Eq.~\eqref{qpS}, giving access to the moments $\langle S_A(t)^n\rangle$ of the entanglement entropy at any time, and to  the out-of-equilibrium evolution of the average entanglement asymmetry.
  }
\end{figure*}

In the present work, we continue the characterization of the entanglement dynamics in the $\nu=1$ QSSEP model. In particular, we analytically derive the full-time probability distribution of the entanglement entropy after a quench from an initial product state. To this end, we adapt the stochastic generalization of the quasiparticle picture developed in Ref.~\cite{alba-25}. As we will show, the crucial ingredient for computing entanglement fluctuations is the inclusion of the statistical correlations between the pairs of entangled quasiparticles, see Fig.~\ref{fig:DQPsketch} for an illustration. This represents a key novelty compared to the case of the average entanglement, or the non-random system, where the pairs are assumed to be uncorrelated. 

This approach also allows us to investigate other interesting aspects of the $\nu=1$ QSSEP model. Here, we further apply it to investigate how particle-number symmetry is restored in the stationary state when the initial configuration breaks it — a problem that has recently attracted significant attention.
To monitor the restoration of symmetry, we employ the entanglement asymmetry, an observable that quantifies the extent to which the symmetry is broken within a subsystem.
Leveraging the diffusive quasiparticle picture, we derive the full-time dynamics of the average asymmetry for generic initial squeezed states that break particle-number symmetry.
An intriguing phenomenon to explore in this context is the quantum Mpemba effect, which occurs when, for a pair of initial states, symmetry is restored faster in the state that initially breaks it more~\cite{amc-23} (see Refs.~\cite{teza-25, acm-25} for a review of the different manifestations of the quantum Mpemba effect and Ref.~\cite{smbtmg-25} for a unifying approach to them).
This particular version of the effect has been observed in various systems — including free and integrable models~\cite{rylands-24, makc-24, carc-24, k-24, rvc-24, yac-24, yca-24, gsb-25, yet-25}, open, monitored, and driven systems~\cite{cma-24, avm-24, dgtm-25, bds-24}, quantum circuits~\cite{lzyz-24, tcd-24, krb-24, fcb-25, ylz-25, amcp-25, yhz-25}, disorder and long-range spin chains~\cite{liu-24-2, yjzxf-25, ya-25, hycmp-25}, and even experimentally~\cite{joshi-24, xu-25} — as well as in the non-random analogue of the $\nu=1$ QSSEP. We determine the microscopic conditions for its occurrence in terms of the density of occupied modes of the initial squeezed states and compare them with the non-random case.

The paper is organized as follows. In Sec.~\ref{sec:model}, we introduce the $\nu=1$ QSSEP. In Sec.~\ref{sec:ent_fluct}, after reviewing the results on the average entanglement entropy reported in Ref.~\cite{alba-25}, we extend the diffusive quasiparticle picture that describes it to include the noise-induced correlations between quasiparticle pairs, and we apply this framework to compute the full counting statistics of the entanglement entropy. In Sec.~\ref{sec:asymmetry}, we use this quasiparticle picture to study the entanglement asymmetry after a quantum quench from particle-symmetry-breaking initial states, and we determine the conditions under which the quantum Mpemba effect occurs in the $\nu=1$ QSSEP model. We conclude in Sec.~\ref{sec:conclusions} with a summary and outlook. We include an appendix that describes the numerical methods used in our calculations. 

\section{$\nu$-QSSEP model}\label{sec:model}
In this section, we present the $\nu$-QSSEP model for $\nu = 1$ as originally introduced in Ref.~\cite{alba-25}. As mentioned in the introduction, this model consists of a one-dimensional chain of $L$ sites with non-interacting spinless fermions and random complex hopping amplitudes that are homogeneous in space but stochastic in time. The dynamics is, therefore, described by the following infinitesimal Hamiltonian $dH_t$ which generates the time evolution,
\begin{equation}
\label{model_dH}
    d H_t = \sqrt{D} \sum_{j=1}^L \left( dW_t c^{\dagger}_j c_{j+1} + h.c.\right),
\end{equation}
where $c_j$ ($c^{\dagger}_j$) are the standard annihilation (creation) operators of a spinless fermion at the site $j$. We consider periodic boundary conditions, $c_{j+L}=c_j$. The hopping amplitude $dW_t$ is a complex It\^o increment satisfying $dW_td W_t =\overline{dW}_t\overline{d W_t}= 0$ and $dW_t\overline{d W_t} = dt$. The constant $D$ tunes the size of the fluctuations of the noise and, as we discuss below, determines the time scale of the time evolution. The density matrix $\rho_t = \ket{\psi_t} \bra{\psi_t}$ that describes the state of the full system thus evolves as 
\begin{equation}\label{eq:time_ev}
\rho_{t+dt} = e^{-i dH_t} \rho_t e^{i dH_t}.
\end{equation}
In the following, we are interested in the out-of-equilibrium dynamics obtained by time evolving a lowly entangled initial state $\ket{\psi_0}$. Since the Hamiltonian generator~\eqref{model_dH} is quadratic in the fermionic operators, if the initial state is Gaussian, i.e. it satisfies Wick theorem, then the time-evolved state~\eqref{eq:time_ev} will remain Gaussian at all times~\cite{peschel-03, pe-09}.

This model has a global $U(1)$ symmetry associated with the conservation of the total particle number, i.e. $[dH_t, Q] = 0$, with $Q=\sum_j c^{\dagger}_j c_j$. First, to study the fluctuations of entanglement, we will consider initial states with a definite number of particles for simplicity. Later, we relax this assumption and analyze the evolution and restoration of this $U(1)$ symmetry when the initial state breaks it. 
Due to the Gaussianity of the state at all times, all the physical information is encoded in the two-point function $G_{nm} = \bra{\psi_t} c^{\dagger}_n c_m\ket{\psi_t}$ when the initial state has a definite number of particles. If the initial state violates the particle number symmetry, then we must further take into account the correlations $F_{nm} = \bra{\psi_t} c_{n} c_{m} \ket{\psi_t}$. All the observables that we will study in the following can be obtained from the knowledge of  $G_{nm}$ and $F_{nm}$, see Refs.~\cite{peschel-03,pe-09} and also Appendix~\ref{appA} for a discussion on the numerical techniques used.

\section{Diffusive quasiparticle picture for the entanglement statistics}\label{sec:ent_fluct}

In this section, we study the fluctuations of entanglement 
under the time evolution described by Eq.~\eqref{eq:time_ev}. 
To this end, we divide the chain in two intervals $A$ and $B$ of length $\ell$  and $L-\ell$, respectively, as illustrated in Fig.~\ref{fig:DQPsketch}. In Ref.~\cite{alba-25}, it was found that the average entanglement entropy between $A$ and $B$ under Eq.~\eqref{eq:time_ev} can be described in terms of diffusively propagating entangled pairs of quasiparticles in the scaling limit $L,\ell,t \to \infty$ with $\Theta\equiv Dt/\ell^2$ finite. Here, we show that such diffusive quasiparticle picture can be properly extended to determine the full counting statistics of the entanglement entropy in that scaling limit.

Let us first review the original result derived in Ref.~\cite{alba-25} for the average entanglement entropy $\langle S_{A}(t)\rangle$. We consider a generic initial state with two-site translational invariance and a definite number of particles, i.e. $F_{nm}=0$, such as the N\'eel and the dimer states. Exploiting the translational invariance of the generator of the time evolution~\eqref{model_dH}, the two-point correlation matrix $G_{nm}$ can be expressed in momentum space in the form
\begin{equation}\label{eq:corr_mat}
    G_{nm} = \frac{1}{L} \sum_{k,q} e^{-ikn+iqm} \tilde{G}_{kq}
\end{equation}
at all times. Due to the two-site translational invariance of the initial configuration, at $t = 0$ we have
\begin{equation}\label{eq:initial_Gkq}
    \tilde{G}_{kq} = \frac{1}{2}f(k) \delta_{k,q} + \frac{1}{2} g(k) \delta_{k,q+\pi},
\end{equation}
with $g(k-\pi) = g(k)^*$.
Let $\mathcal{F}(G_A)$ be a generic analytic function of the correlation matrix restricted to the subsystem $A$, i.e. $(G_A)_{m, n}=G_{m, n}$, $m, n\in A$. Using the stationary phase approximation~\cite{cef-12}, it can be shown that, for an infinite chain (i.e., in the thermodynamic limit $L\to\infty$) in the diffusive scaling limit in which  $\ell,t \to \infty$ and $\Theta=Dt/\ell^2$ is finite, 
the expectation value of the trace of $\mathcal{F}(G_A)$ is given by \cite{alba-25}
\begin{multline}
\label{expTrF}
   \langle{\rm Tr}\left[\mathcal{F}(G_A)\right]\rangle=
   \frac{1}{2}\int_{-\infty}^\infty d\xi P(\xi)\int\frac{d k}{2\pi}\left[\max(\ell-2|\xi|, 0)\left[\mathcal{F}(\lambda_+)+\mathcal{F}(\lambda_-)\right]\right.\\\left.+\min(2|\xi|, \ell)\left[\mathcal{F}(f(k)/2)+\mathcal{F}(f(k+\pi)/2)\right]\right],
\end{multline}
where $\lambda_{\pm}$ are the eigenvalues of the matrix
\begin{equation}
    Q = \begin{pmatrix}
        f(k+\pi) & g(k+\pi) \\
        g(k) & f(k)
    \end{pmatrix},
\end{equation}
and $P(\xi,t) = e^{-\xi^2/4Dt}/\sqrt{4 \pi Dt}$ is the probability distribution function of a Gaussian random variable with zero mean and variance $2 Dt$. If we now specialize the formula~\eqref{expTrF} to the entanglement entropy $ S_A = -\text{Tr}[\rho_A \log \rho_A]$, then we find the following time evolution for its average value,
\begin{equation}
\label{avgSdqp}
    \langle S_A(t)\rangle = s \, \ell \int d\xi P(\xi,t)\, \text{min}(2|\xi|/\ell,1),
\end{equation}
where
\begin{equation}
\label{s1}
    s = - \int \frac{dk}{2\pi} \left[\frac{f(k)}{2}\log \frac{f(k)}{2} + \left(1-\frac{f(k)}{2} \right) \log \left( 1- \frac{f(k)}{2}\right)
    \right].
\end{equation}
According to Eq.~\eqref{avgSdqp}, the dynamics of the average entanglement entropy under Eq.~\eqref{eq:time_ev} is characterized by an initial $\sqrt{t}$ growth followed by a saturation, at a time $t \gtrsim \ell^2$, to the same value predicted by the generalized Gibbs ensemble (GGE) in the case of non-random hopping amplitudes~\cite{fc-08, ac-18}.

As already mentioned, the results in Eqs.~\eqref{expTrF} and~\eqref{avgSdqp} admit a quasiparticle interpretation. At $t = 0$, pairs of entangled quasiparticles are produced at each spatial point of the system and then they travel diffusively, as shown in Fig.~\ref{fig:DQPsketch}. The distance between the two quasiparticles of a pair is $2 |\xi|$ with a probability determined by the distribution $P(\xi,t)$. Since for each noise realization all the entangled pairs perform the same random walk, the number of pairs that are fully contained within the subsystem $A$ is $\max(\ell-2|\xi|, 0)$, whereas the number of pairs with one quasiparticle in region $A$ and the other in $B$ is $\min(2|\xi|, \ell)$. 
Therefore, the first term in Eq.~\eqref{expTrF} is the contribution from the  pairs that are fully contained  inside $A$ and the second one corresponds to the contribution of the pairs that are shared between $A$ and the rest. The contribution of each pair is the same as in the non-random model. In particular,
the entanglement between $A$ and $B$ is given according to Eq.~\eqref{avgSdqp} by the number of quasiparticles shared between these subsystems, i.e. $\text{min}(2 |\xi|,\ell)$, multiplied by their contribution~\eqref{s1}. Notice that, unlike in the case of non-random hoppings, the dependence of the distance $|\xi|$ on the momentum $k$ of the quasiparticles is completely washed away as a consequence of the average over the noise. However, we will show that this will not be longer true when considering averages of functions of the correlation matrix such as $\langle \text{Tr}[\mathcal{F}_1(G_A)]\, \text{Tr}[\mathcal{F}_2(G_A)] \cdots \rangle$ or more generically $\langle h(\text{Tr}[\mathcal{F}_1(G_A)],\text{Tr}[\mathcal{F}_2(G_A)],\cdots) \rangle$, with $h$ an analytic function of multiple variables. The relevance of considering such expectation values relies on the fact that their knowledge is crucial to the determination of the higher moments of the entanglement entropy $\langle S_A(t)^n\rangle$, and the presence of correlations between quasiparticles with different momentum will be the key point in the generalization of Eq.~\eqref{avgSdqp}.

The existence of a quasiparticle picture is hinted at by the presence of well-defined, stable excitations at all times, regardless of the stochasticity of the hopping amplitudes $dW_t$. The infinitesimal generator of the dynamics~\eqref{model_dH} takes in momentum space the simpler form
\begin{equation}
\label{dhmomspace}
    dH_t = \sum_k d\varepsilon_{t,k} c^{\dagger}_k c_k,
\end{equation}
where $d \varepsilon_{t,k} = \sqrt{D}(dW_t e^{i k} + \overline{dW}_t e^{-ik}) = 2 \sqrt{D} \,\text{Re}\{dW_t e^{i k}\}$ are real It\^o increments that satisfy $d\varepsilon_{t,k}d\varepsilon_{t,q} = 2 D dt \cos(k-q)$. The crucial point is that infinitesimal generators at different times commute. Therefore, we can think of the excitations with momentum $k$ as having a stochastic energy $\varepsilon_k$, whose probability distribution is time-dependent. This energy is given by $\varepsilon_k (t)= \sqrt{D}(W_t \,e^{ik}+ \overline{W}_t \,e^{-ik}) = 2 \sqrt{D}\, \text{Re}\{W_t\,e^{ik}\}$, where $W_t$ is the random variable $W_t = W_{R,t} + i W_{I,t}$ and $W_{R(I),t}$ describe independent real Brownian motions with zero mean and variance $Dt$.

Based on this idea, to generalize the standard quasiparticle picture of the non-random model~\cite{ac-18}, we observe that the distance $\xi_k$ traveled by a quasiparticle with momentum $k$ is determined by the stochastic energy $\varepsilon_k$. Indeed, the infinitesimal distance is obtained from
\begin{equation}
\label{eq:xi-def}
\xi_k(t) = \partial_k \varepsilon_k(t) = 2 \sqrt{D}\, \text{Re}\{ W_t \,e^{ik}\}.
\end{equation}
Note that the factor $i$ coming from the derivative $\partial_k$ has been absorbed in $W_t$, as its probability distribution remains invariant.
Now, by considering a finite time interval, the probability distribution of $\xi_k$ is independent on $k$ and it is given by $P(\xi,t)$ as in Eq.~\eqref{expTrF}, since $W_t$ is a complex Brownian motion. We anticipate that this, in turn, implies that for observables linear in $\xi_k$, the dependence on $k$ drops out.  On the other hand, $\xi_k=2\sqrt{D}\text{Re}\{W_te^{ik}\}$ and $\xi_q=2\sqrt{D}\text{Re}\{W_t e^{iq}\}$ are not independent, as follows from their definition.
Let us see how to account for these statistical correlations between quasiparticles.
We consider again a generic analytic function  $\mathcal{F}(G_A)$ of the subsystem correlation matrix $G_A$. From Eq.~\eqref{expTrF}, we first define the function $ z_{\mathcal{F}}(W_t)$ 
\begin{multline}\label{eq:z_F}
    z_{\mathcal{F}}(W_t) = \frac{1}{2} \int \frac{dk}{2 \pi}\left[\max(\ell-2|\xi_k|, 0)\left[\mathcal{F}(\lambda_+)+\mathcal{F}(\lambda_-)\right]\right.\\ \left.+\min(2|\xi_k|, \ell)\left[\mathcal{F}(f(k)/2)+\mathcal{F}(f(k+\pi)/2)\right]\right].
\end{multline}
Here $\lambda_{\pm}$ and $f(k)$ are the same  as in Eq.~\eqref{expTrF}, but we have replaced $\xi$ by $\xi_k=\partial_k\varepsilon_k(t)=2\sqrt{D}\mathrm{Re}\{W_t e^{ik}\}$. 
We conjecture that the average value of a generic analytical function $h(\cdot,\cdot,\dots)$ of traces of different functions $\mathcal{F}_j(G_A)$, $j=1,2,\dots$, is given by 
\begin{equation}
\label{generaldqp}
    \langle h(\text{Tr}[\mathcal{F}_1(G_A)],\text{Tr}[\mathcal{F}_2(G_A)],\cdots)\rangle \underset{L,\ell,t \to \infty}{=} \langle h(z_{\mathcal{F}_1},z_{\mathcal{F}_2},\cdots)\rangle_{\Theta},
\end{equation}
where we have introduced $\langle\cdot \rangle_{\Theta}$ as a shorthand notation for the average with respect to the probability distribution of the complex Brownian motion $W_{\Theta = Dt/\ell^2}$. If $h(x) = x$, then we recover the result of Eq.~\eqref{expTrF}.

To understand how the general ansatz in Eq.~\eqref{generaldqp} works, we can specialize ourselves to the entanglement entropy. As we will show, as a consequence of Eq.~\eqref{generaldqp}, it is possible to formally write, in the diffusive scaling limit, the time evolution of the $n-$Rényi entropy $ S_A^{(n)} = 1/(1-n) \log \text{Tr}[\rho_A^{n}]$ --- and also of the von Neumann entropy in the limit $n\to 1$ --- as
\begin{equation}
\label{qpS}
    S^{(n)}_A(\Theta) \underset{L,\ell,t \to \infty}{\asymp}  \int \frac{dk}{2 \pi} \text{min}(2|\xi_k(\Theta)|,\ell)\,s_n(k),
\end{equation}
where the symbol $\asymp$ is understood as an equality inside expectation values $\langle\cdot\rangle$. 
The factor $s_n(k)$ is the contribution of the mode $k$ to the $n$-Rényi entropy, which is the same as in the non-random case~\cite{ac-18}, i.e. $s_n(k) = \log(n_k^n+(1-n_k)^n)/(1-n)$, with $n_k$ the density of occupied modes $n_k=\bra{\psi_0}c_k^\dagger c_k\ket{\psi_0}=f(k)/2$. 
To connect with Eq.~\eqref{generaldqp}, we recall that the $n-$R\'enyi entanglement entropy is expressed in terms of the two-point correlation matrix $G_A$ via~\cite{peschel-03}
\begin{equation}
S_A^{(n)}=\frac{1}{1-n}{\rm Tr}\log\left[G_A^n+(I-G_A)^n\right].
\end{equation}
Thus Eq.~\eqref{qpS} corresponds to the 
particular choice $h(x)=x$ and $\mathcal{F}(x) = \log(x^n+(1-x)^n)/(1-n)$ in 
Eq.~\eqref{generaldqp}.
Taking the expectation value of Eq.~\eqref{qpS}, the dependence on $k$ is completely washed away in $\langle |\xi_k(t)|\rangle$, since $\xi_k$ is distributed according to $P(\xi, t)$, which is independent of $k$. Using this property, we immediately obtain 
 \begin{equation}
 \label{avgS}
     \langle S^{(n)}_A(\Theta) \rangle = \left( \int \frac{dk}{2 \pi} s_n(k)\right) \langle \text{min}(2|\xi|,\ell) \rangle_{\Theta},
 \end{equation}
 This result is exactly equivalent to Eq.~\eqref{avgSdqp}, obtained \textit{ab initio} in Ref.~\cite{alba-25} through the stationary phase approximation.
 
Let us now show how Eq.~\eqref{qpS} can also predict the time evolution of higher moments $\langle (S^{(n)}_A(t))^m\rangle$.
We first focus on the variance of the von Neumann entropy $\sigma_S^2(t) = \langle S_A(t)^2 \rangle- \langle S_A(t)\rangle^2$, although all the following arguments are valid for generic Rényi index. 
We directly apply Eq.~\eqref{qpS} in the expectation value $\langle S_A(t)^2\rangle$, which is equivalent to Eq.~\eqref{generaldqp} for $h(x) = x^2$ and $\mathcal{F}(x) = \log(x^n+(1-x)^n)/(1-n)$. Then we find that, in the limit $L,\ell,t \to \infty$ with $\Theta$ finite,
\begin{equation}
\label{s2dqp}
\begin{split}
    \langle S_A(\Theta)^2 \rangle &= \ell^2 \int \frac{dk\,dk'}{(2 \pi)^2} s(k) s(k') \langle\text{min}\left(2 |\xi_k (\Theta)|/\ell,1\right)\text{min}\left(2 |\xi_{k'}(\Theta)|/\ell,1\right) \rangle_{\Theta}  \\
    &= \ell^2 \int \frac{dk\,dk'}{(2 \pi)^2} s(k) s(k') \left\langle\text{min}\left(\frac{4}{\ell} |\text{Re}\{W_\Theta \,e^{ik}\} |,1\right)\text{min}\left(\frac{4}{\ell} |\text{Re}\{W_\Theta \,e^{ik'}\} |,1\right)\right\rangle_{\Theta},
\end{split}
\end{equation}
where in the second line the average $\langle\cdot \rangle_{\Theta}$ is calculated with respect to the probability distribution of the normally distributed complex variable $W_\Theta$.
We were not able to compute analytically this expectation value. However, we can numerically estimate it by sampling over $W_\Theta$ and averaging the result of many realizations. We then compare this result with the variance obtained by sampling from the exact evolution of the correlation matrix $G$ under Eq.~\eqref{eq:time_ev}, as explained in Appendix~\ref{appA}. In Fig.~\ref{fig:VarianceS}, we show this comparison in a quench from the N\'eel state, for which $s(k) = \log 2$ for all $k$. We observe that Eq.~\eqref{s2dqp}, derived from the diffusive quasiparticle picture, correctly reproduces the time evolution of the variance of the entanglement entropy.

\begin{figure*}[t]
\centering
\includegraphics[width=0.6\textwidth]{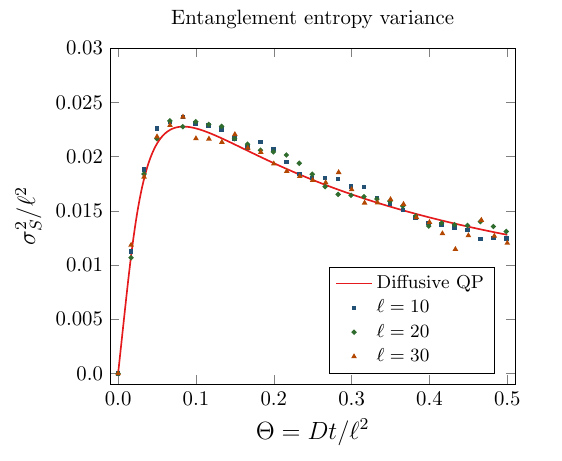}
  \caption{\label{fig:VarianceS} Time evolution of the rescaled variance of the entanglement entropy $\sigma_S^2/\ell^2$ under the dynamics in Eq.~\eqref{model_dH} starting from the Néel state, for different subsystem sizes $\ell$ in the thermodynamic limit $L\to\infty$. The symbols correspond to the variance of the exact entanglement entropy, computed from the time-evolved correlation matrix as explained in Appendix~\ref{appA}, for 1000 realizations. The red curve is the prediction in Eq.~\eqref{s2dqp} obtained from the diffusive quasiparticle picture.
  }
\end{figure*}
\begin{figure*}[t]
\centering
\includegraphics[width=0.7\textwidth]{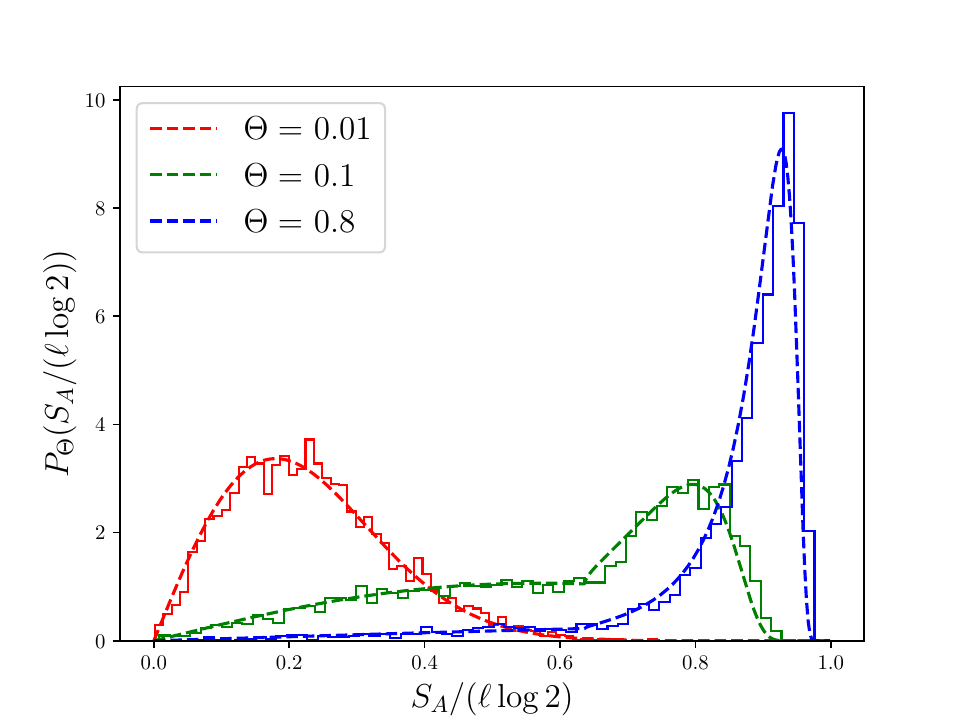}
  \caption{\label{fig:pdfS}
  Out-of-equilibrium probability distribution function of the entanglement entropy at different times $\Theta = D t/\ell^2$ after a quench with the Hamiltonian \eqref{model_dH}, starting from the Néel state and taking a subsystem of length $\ell=30$ and $L\to\infty$.
  The entanglement entropy $S_A$ is rescaled by its asymptotic value $\ell \log2$ in the stationary state.
  The dashed curves represent the prediction~\eqref{eq:prob_dist} of the diffusive quasiparticle picture. The histograms are the result of sampling the exact entanglement entropy, computed numerically from the time-evolved two-point correlation matrix for different noise realizations. The number of samples for each time considered is $\sim 10^4$. 
  }
\end{figure*}

Using the same procedure, it is straightforward to construct the $m$-th moment of $S_A(t)$ in terms of the $m$-point function of $\min(2|\xi_k|, \ell)$, which counts the number of quasiparticles shared between $A$ and $B$. As a consequence, this picture should also apply to determining the full probability distribution $P_\Theta(S_A)$ of the entanglement entropy. This distribution can be obtained from the moment generating function $\chi_\Theta(\lambda)=\langle e^{\lambda S_A(\Theta)}\rangle$ as
\begin{equation}\label{eq:prob_dist}
P_\Theta(S_A)=\int d\lambda e^{-\lambda S_A}
\chi_\Theta(\lambda).
\end{equation}
The ansatz proposed in Eq.~\eqref{qpS} predicts the following expression for the generating function $\chi_\Theta(\lambda)$
\begin{equation}
\chi_\Theta(\lambda)=\biggl< \exp\left(\lambda\int \frac{dk}{2\pi}\min(2|\xi_k(\Theta)|, \ell)s_1(k)\right)\biggr>_\Theta,
\end{equation}
from which we can determine numerically the time evolution of probability distribution $P_{\Theta}(S_A)$. In Fig.~\ref{fig:pdfS}, we compare this prediction with the statistics of entanglement obtained from the exact time-evolved correlation matrix for different noise realizations, finding an excellent agreement. 
In both Figs.~\ref{fig:VarianceS} and~\ref{fig:pdfS}, we can see that the entanglement fluctuations rapidly grow at early times and then they slowly decay as $1/\sqrt{\Theta}$. Observe that in the transition between these two regimes, i.e. around $\Theta\sim0.1$, the distribution probability $P_\Theta(S_A)$ presents a non-analiticity due to the quasiparticle counting function $\min(2|\xi_k|,\ell)$. The fluctuations of
the entanglement entropy vanish at late times. In 
fact, in the limit $\Theta\to\infty$, the 
entanglement probability distribution~\eqref{eq:prob_dist}  tends to $P_{\Theta\to\infty}(S_A)=\delta(S_A-S_A^*)$, where $S_A^*$ is equal to the stationary entanglement entropy of the non-random model. In contrast, in the fully inhomogeneous QSSEP ($\nu\to\infty$), the probability distribution of the entanglement entropy satisfies a large deviation principle at large times, as shown in Ref.~\cite{bp-21}. It would be interesting to analyze what happens for $\nu>1$ but finite.

\subsection{Some checks via the stationary phase approximation}

Here we provide arguments supporting the non-trivial ansatz proposed in Eq.~\eqref{generaldqp} by leveraging the standard tool of the stationary phase approximation, which allows for \emph{ab initio} calculations~\cite{cef-12}. Specifically, we first consider $\langle \text{Tr}[G_A^2] ^2\rangle$ and then the generic product $\langle \text{Tr}[G_A^2] ^m\rangle$. We show that these quantities can be computed using the stationary phase approximation and that the structure expected from the diffusive quasiparticle picture in Eq.~\eqref{generaldqp} naturally emerges in these examples.

For simplicity, we take as initial configuration the Néel state. In this case, we have $f(k)=g(k)=1$ in Eq.~\eqref{eq:initial_Gkq}. Specializing Eq.~\eqref{expTrF} to $\mathcal{F}(G_A)={\rm Tr}(G_A^2)$, our conjecture reads as 
\begin{equation}\label{eq:ansatz_GA2}
{\rm Tr}[G_A^2(\Theta)]\underset{L,\ell,t \to \infty}{\asymp} \frac{\ell}{2}\int\frac{dk}{2\pi}\left(1+x_{\xi_k}(\Theta)\right),
\end{equation}
and, applying Eq.~\eqref{generaldqp}, we expect
\begin{equation}
\label{trtrdqp}
    \langle \text{Tr}[G_A^2(\Theta)] \text{Tr}[G_A^2(\Theta)] \rangle = \left(\frac{\ell}{4}\right)^2 \left(1 + 2 \langle x_{\xi} \rangle + \int \frac{dk \, dk'}{(2 \pi)^2} \langle x_{\xi_k}x_{\xi_{k'}} \rangle \right),
\end{equation}
where $x_{\xi_k}$ is defined as 
\begin{equation}
\label{eq:x-def}
x_{\xi_k} = \text{max}(1-2|\xi_k(\Theta)|/\ell,0). 
\end{equation}
Notice that $x_{\xi_k}$  is the number of pairs of quasiparticles at relative distance $2|\xi_k|$, which are fully contained in $A$.

Let us see how this result can be also derived from the stationary phase approximation. We begin by rewriting $\langle \text{Tr}[G_A^2] ^2\rangle$ using
the decomposition in momentum space~\eqref{eq:corr_mat} of $G_A$ and the identity
\begin{equation}
    \sum_{n=1}^\ell e^{ikn} = \frac{\ell}{4} \int_{-1}^1 d\xi \,w(k) e^{i(\ell \xi +\ell +1)k/2},
\end{equation}
where $w(k) = k/\sin(k/2)$. We obtain
\begin{multline}
\label{exptrtr}
    \langle \text{Tr}[G_A^2(t)] \text{Tr}[G_A^2(t)]\rangle = \left(\frac{\ell}{4}\right)^4 \int _{-1}^{1}\left(\prod _j d \xi_j\right) \int \left(\prod _j \frac{d k_j\,d q_j}{(2 \pi)^2}\right) w_{12}^2 w_{34}^2\\ \times e^{- i \ell (k_1-q_2)\xi_1/2 - i \ell (k_2-q_1)\xi_2/2 - i \ell (k_3-q_4)\xi_3/2 - i \ell (k_4-q_3)\xi_4/2}\left\langle\prod_{j=1}^4 G_{k_j,q_j}(t) \right\rangle,
\end{multline}
where we have defined $w_{ij}=w(k_i-k_j)$.
As shown in Ref.~\cite{alba-25}, the dynamics of the expectation value of the product of two-point functions in momentum space is governed by 
\begin{equation}
\label{defene}
    d\left\langle\prod_{j=1}^4 G_{k_j,q_j}(t) \right\rangle = \varepsilon({\{k_j,q_j\}}) \left\langle\prod_{j=1}^4 G_{k_j,q_j}(t) \right\rangle dt.
\end{equation}
The energy $\varepsilon({\{k_j,q_j\}})$ is zero if $q_j = k_j$ for any $j$. This choice of momenta corresponds precisely to the stationary state value of the correlator.  For other choices of the momenta, the energy $\varepsilon({\{k_j,q_j\}})$ is generally negative, implying an exponential decay in time of the expectation value in Eq.~\eqref{defene}. We will be interested in the leading contribution to Eq.~\eqref{exptrtr}, apart from the stationary state value, in the scaling limit $\ell,t \to \infty$. Following Ref.~\cite{alba-25}, we observe that the leading term in the scaling limit is obtained from correlators with \emph{paired} momenta. We define two momenta $k_j$ and $k_{j'}$ as being paired if $q_j = \pm k_j$ and $q_{j'} = \mp k_{j'}$, and we will refer to this pairing using the notation $(j\,j')$. The energy corresponding to $p\geq1$ paired momenta is given by
\begin{equation}
\label{epsp}
    \varepsilon_p(\{k_j\}) = -8D\left(p + \sum_{\text{paired} \,j,j'} \cos(k_j-k_{j'})\right).
\end{equation}
In the case of expectation values of the form $\langle \text{Tr}[\mathcal{F}(G_A)]\rangle$ studied in Ref.~\cite{alba-25}, the leading contribution corresponds to having an even number of paired momenta with all the remaining unpaired momenta satisfying $q_j = k_j$.  However, when considering more complex expectation values such as in Eq.~\eqref{exptrtr}, the presence of multiple traces imposes further constraints on the pairings that give the leading contribution. 

Let us analyze all the possible pairings in Eq.~\eqref{exptrtr}. We begin with no pairings, i.e. $q_j = k_j$ for any $j$. In this case, $\varepsilon_0 = 0$ and the terms coming from the two traces in Eq.~\eqref{exptrtr} completely decouple. Their contribution is thus equal to the stationary value of $\langle \text{Tr}[G_A^2]\rangle$ squared, which is $C_0 = (\ell/4)^2$. In the case of a single pairing $(j\,j')$, whose contribution is denoted by $C_{(j\,j')}$, there are two possibilities: either both $j$ and $j'$ are indices associated with the same trace, such as $(1\,2)$ and $(3\,4)$, or $j$ and $j'$ come from a different trace, for instance $(1\,3)$ and $(2\,4)$. In the first scenario, due to the form of Eq.~\eqref{defene}, the two traces completely decouple again leaving us with the computation of the product of $\langle \text{Tr}[G_A^2]\rangle$ for zero pairing and of $\langle \text{Tr}[G_A^2]\rangle$ for one pairing. The case of zero pairing has just been discussed above while the computation for one pairing was done in Ref.~\cite{alba-25}; the final result is $C_{(1\,2)}=C_{(3\,4)}=(\ell/4)^2 \langle x_{\xi}\rangle$.
In the second scenario, we can consider the pairing $(1\,3)$ (other combinations of the same type, i.e., with indices $j$ and $j'$ in different traces give the same result). Its contribution to the expectation value in Eq.~\eqref{exptrtr} is
\begin{multline}\label{eq:C_13}
    C_{(1\, 3)} = \left(\frac{\ell}{8}\right)^4 \int _{-1}^{1}\left(\prod _j d \xi_j\right) \int \left(\prod _j \frac{d k_j}{2 \pi}\right) w_{12}^2 w_{34}^2 \,e^{-8Dt(1+\cos(k_1-k_3))}\\ \times e^{- i \ell (k_1-k_2)\xi_1/2 - i \ell (k_2-k_1-\pi)\xi_2/2 - i \ell (k_3-k_4)\xi_3/2 - i \ell (k_4-k_3+\pi)\xi_4/2} ,
\end{multline}
    where we have used Eqs.~\eqref{defene} and \eqref{epsp} as well as the specific form~\eqref{eq:initial_Gkq} of the two-point correlation matrix at $t=0$ for the N\'eel state. Because of the cyclicity  of the momenta $k_j$ in the exponent in Eq.~\eqref{eq:C_13}, the terms $e^{\pm i\pi\ell\xi_{2(4)}/2}$ cannot simply be eliminated by redefining the $k_j$'s. For this reason, if one performs the change of coordinates $\zeta_1 = \xi_2-\xi_1,\, \zeta_2 =\xi_2$ and $\zeta_3 = \xi_4-\xi_3,\, \zeta_4 =\xi_4$, the variables $\zeta_2$ and $\zeta_4$ remain in the integrand as $e^{\pm i \pi \ell \zeta_{2(4)}/2}$. Each integral in $\zeta_2,\zeta_4$ gives then an extra $O(\ell^{-1})$ factor, which is subleading. Therefore, the contribution of pairings $(j \, j')$ where $j$ and $j'$ correspond to independent traces is subleading and it can be neglected in the diffusive scaling limit. The last term we must consider comes from having two pairings, each corresponding to one of the two traces, i.e. $(1\,2)(3\,4)$. Applying Eqs.~\eqref{defene}~and~\eqref{epsp}, we find that its contribution is of the form
\begin{multline}\label{eq:C1234}
   C_{( 1\,2)(3\,4)} = \left(\frac{\ell}{8}\right)^4 \int _{-1}^{1}\left(\prod_j d \xi_j\right) \int \left(\prod _j \frac{d k_j}{(2 \pi)} \right) w_{12}^2 w_{34}^2 \, e^{-8\gamma t(2 + \sum_{j> j'} \cos(k_j-k_{j'}))} \\ \times e^{- i \ell (k_1-k_2+\pi)\xi_1/2 - i \ell (k_2-k_1-\pi)\xi_2/2 - i \ell (k_3-k_4+\pi)\xi_3/2 - i \ell (k_4-k_3-\pi)\xi_4/2}.
\end{multline}
The $\pi$-shifts in the exponent can, this time, be removed by redefining $k_{1(3)}+\pi\mapsto k_{1(3)}$. In terms of the relative coordinates $\zeta_j$, Eq.~\eqref{eq:C1234} takes the form
\begin{multline}
\label{statph}
    C_{( 1\,2)(3\,4)} = \left(\frac{\ell}{8}\right)^4 \int _{-2}^{2} d\zeta_2 d \zeta_4\, \mu(\zeta_2) \mu(\zeta_4) \int \left(\prod _j \frac{d k_j}{(2 \pi)} \right) w^2_{12} w^2_{34}
    e^{- i\ell(k_2-k_1)\zeta_2/2 - i \ell(k_4-k_3)\zeta_4/2} \\e^{-8\gamma t (2-\cos(k_1-k_2)-\cos(k_3-k_4) + \cos(k_1-k_3)+\cos(k_2-k_4)-\cos(k_1-k_4)-\cos(k_2-k_4))},
\end{multline}
where $\mu(\zeta) = \text{max}(0,\text{min}(1,1-\zeta)+\text{min}(1,1+\zeta))$ are the result of the integration of $\zeta_1$ and $\zeta_3$. Since the phase in the integrand of Eq.~\eqref{statph} rapidly oscillates when $\ell\to\infty$, we can obtain its asymptotic
behavior applying the stationary phase approximation. 
The stationarity condition for $\zeta_2$ and $\zeta_4$ in the limit $\ell,t \to \infty$ implies that we can expand the integrand around $k_2 \simeq k_1$ and $k_4 \simeq k_3$. Expanding up to the second order in that limit allows us to approximate the integrals over $k_2,k_4$ as Gaussian integrals and apply the standard formulas for them. The result, as a function of the scaling variable $\Theta$, is
\begin{equation}
\label{trtrsq}
    C_{( 1\,2)(3\,4)} =\left(\frac{\ell}{4}\right)^2 \int \frac{dk\,dk'}{(2 \pi)^2} \bigg[\frac{1}{2 \pi \, 2^4} \int d\zeta d\zeta' \mu(\zeta) \mu(\zeta') \frac{e^{-\frac{1}{64 \Theta} \frac{\zeta^2 + \zeta'^2-2 \zeta \zeta'\cos(k-k')}{\sin(k-k')}}}{8 \Theta |\sin(k-k')|} \bigg],
\end{equation}
where $\mu(\zeta)$ is the same as in~\eqref{statph}.
Notice that the term inside the square bracket can be interpreted as the expectation value of $\mu(\zeta)\mu(\zeta')/4$ with respect to the probability distribution $P_{\bm{\zeta},\bm{k}} \equiv P(\zeta,\zeta';k,k')$ of two correlated Gaussian random variables satisfying $\mathbb{E}_{P_{\bm{\zeta},\bm{k}}}[\zeta \,\zeta'] = 2 \Theta\cos(k-k')$. 

Finally, if we sum all the leading contributions $C_p$, we obtain that
\begin{equation}
\label{statphtrtr}
    \langle \text{Tr}[G_A^2] \text{Tr}[G_A^2]\rangle  = \left( \frac{\ell}{4}\right)^2\left(1+ 2 \langle x_{\xi} \rangle + \int \frac{dk\,dk'}{(2\pi)^2} \frac{1}{4} \mathbb{E}_{P_{\bm{\zeta},\bm{k}}}[ \mu(\zeta) \mu(\zeta')]\right).
\end{equation}
The factor $2$ in front of $\langle x_{\xi}\rangle$ comes from having two terms with one pairing of momenta. 

\begin{figure*}[t]
\centering
\includegraphics[width=0.6\textwidth]{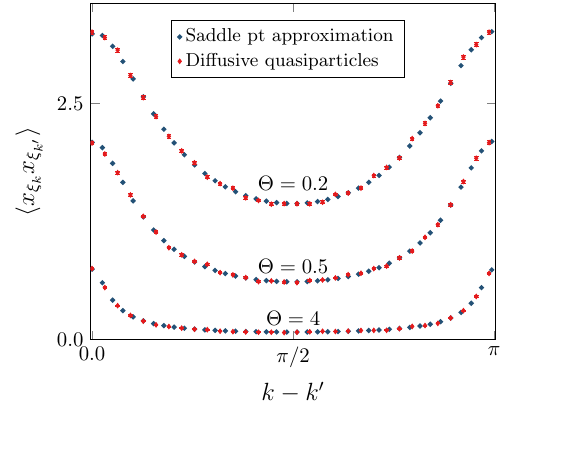}
  \caption{\label{fig:2ptfun}Numerical check of the identity in Eq.~\eqref{ident} for different values of the rescaled time $\Theta=Dt/\ell^2$. The blue circles correspond to the right-hand side of Eq.~\eqref{ident} obtained by evaluating numerically the integral in the square bracket of Eq.~\eqref{trtrsq}. The red diamonds are the value obtained for the average $\langle x_{\xi_k} x_{\xi_{k'}} \rangle = \langle x_{\xi_0} x_{\xi_{k'-k}} \rangle$ by sampling over $4\cdot10^4$ noise realizations.
  }
\end{figure*}

Equation~\eqref{statphtrtr} must be compared with the ansatz in Eq.~\eqref{trtrdqp}. In Fig.~\ref{fig:2ptfun}, we perform the comparison focusing on the last term in~\eqref{statphtrtr}, which is the nontrivial one. By comparing~\eqref{statphtrtr} and~\eqref{trtrdqp} one obtains that the following identity holds
\begin{equation}
\label{ident}
\mathbb{E}_{P_{\bm{\zeta},\bm{k}}}[ \mu(\zeta) \mu(\zeta')] =  4\langle x_{\xi_k} x_{\xi_{k'}}\rangle,
\end{equation}
where in the right-hand side $x_{\xi_k}=\max(1-2|\xi_k(\Theta)|,0)$ and the average is over different realizations of the Brownian motion at different times $\Theta$. 
Fig.~\ref{fig:2ptfun} supports Eq.~\eqref{statphtrtr}. 

\begin{figure*}[t]
\centering
\includegraphics[width=0.6\textwidth]{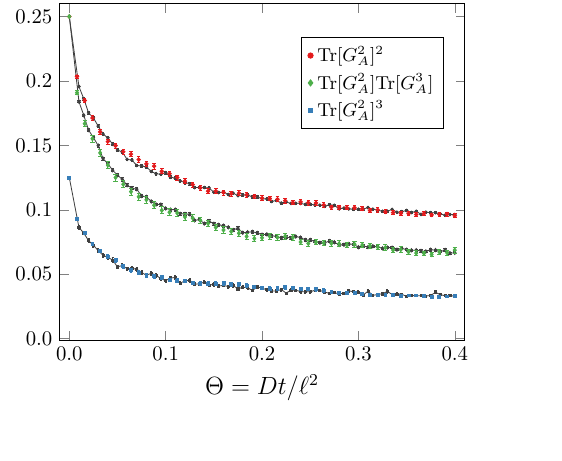}
  \caption{\label{fig:TrTr} Time evolution of the average of different products of traces of the reduced correlation matrix $G_A$ under the dynamics in Eq.~\eqref{model_dH} starting from the Néel state. The red, green, and blue symbols are the average obtained by exactly evolving the Néel state for different realizations of the noise. We take a subsystem of size $\ell = 20$ and $L\to \infty$. The gray symbols joined by a line correspond to the predictions~\eqref{generaldqp} of the diffusive quasiparticle picture. The higher point functions of $x_{\xi_k}$ that appear when applying Eq.~\eqref{generaldqp} are estimated by numerical sampling. In both cases, the error bars are the standard deviation of the mean over the samples considered.
  }
\end{figure*}

We can further support the ansatz~\eqref{generaldqp} by considering the  expectation value of the form $\langle\text{Tr}[G_A^2]^m \rangle$ for generic integer $m>0$. By using Eq.~\eqref{eq:ansatz_GA2}, the diffusive quasiparticle picture predicts
\begin{equation}
\label{trG2mdqp}
    \langle\text{Tr}[G_A^2]^m \rangle = \left(\frac{\ell}{4} \right)^m \left\langle \left(1+ \int\frac{dk}{2\pi} x_{\xi_k}\right)^m\right\rangle.
\end{equation}
One can employ the stationary approximation to compute $\langle\text{Tr}[G_A^2]^m \rangle$, as we have just done for $\langle\text{Tr}[G_A^2]^2 \rangle$. One obtains a sum of contributions given by all the possible ways of pairing the quasimomenta $k_j$. To obtain a nonzero contribution, again, paired momenta have to be in the same trace. Therefore, even though the explicit computation of the leading contribution for a generic pairing choice is rather complicated, by counting the multiplicity of pairings, one obtains 
\begin{equation}
\label{trG2m}
    \langle\text{Tr}[G_A^2]^m \rangle = \left(\frac{\ell}{4} \right)^m \left(1+m C_1 + \binom{m}{2} C_2 + \cdots + C_m\right),
\end{equation}
where the first term corresponds to no pairings ($q_j = k_j$), and $C_p$ is the contribution with $p$ pairings. We also know that 
\begin{equation}
C_1 = \langle x_{\xi}\rangle,\quad  C_2 = \int \frac{dk\,dk'}{(2\pi)^2} \frac{1}{4} \mathbb{E}_{P_{\bm{\zeta},\bm{k}}}[ \mu(\zeta) \mu(\zeta')],
\end{equation}
as explicitly computed for $m=2$.  It is clear that the expansion~\eqref{trG2m} has the same structure as the binomial expansion of the ansatz in Eq.~\eqref{trG2mdqp}, in which each term can be identified with the contribution $C_p$ of $p$-pairings. In Fig.~\ref{fig:TrTr}, we numerically check the validity of Eq.~\eqref{trG2mdqp} for $m = 2,3$. The advantage of Eq.~\eqref{trG2mdqp} is that it allows one to directly compute the contribution $C_p$ from  multiple pairings, without resorting to stationary phase techniques. In this way, one can quickly derive the dynamics of more complex quantities, such as the higher moments of the entanglement entropy, as discussed earlier, or the entanglement asymmetry as shown in the next section.

To conclude, in Fig.~\ref{fig:TrTr}, we provide additional numerical evidence of the validity of Eq.~\eqref{generaldqp} by comparing its prediction against the exact dynamics for an expectation value of the form $\langle {\rm Tr}[G_A^n]^m{\rm Tr}[G_A^{n'}]^{m'}\rangle$. 

\section{Dynamical symmetry restoration and quantum Mpemba effect}\label{sec:asymmetry}

The diffusive quasiparticle picture derived in the previous section allows us to give predictions for the time evolution of the statistics of quite generic functions of $G_A$. In this section, we employ it to study the dynamics of the entanglement asymmetry. Our goal is to characterize, at least in the $\nu=1$ QSSEP, the local dynamical restoration of a symmetry broken by the initial state and the emergence of the quantum Mpemba effect.

The entanglement asymmetry quantifies the extent to which a symmetry transformation generated by a charge $Q_A$ is broken within a subsystem $A$. In this work, we will restrict ourselves to the $U(1)$ group of transformations generated by the fermionic particle number. Therefore, we focus on the entanglement asymmetry for an Abelian charge referring to, e.g., Refs.~\cite{cv-24, rac-24} for its generalization to non-Abelian transformation groups, and to Refs.~\cite{benini-25, akw-25} for generalized symmetries. To define the entanglement asymmetry, it is necessary to introduce the symmetrized reduced density matrix $\rho_{A,Q} = \sum_q \Pi_q \rho_A \Pi_q$, where $\Pi_q$ is the projector onto the eigenspace of $Q_A$ corresponding to the eigenvalue $q$. By construction, $[\rho_{A, Q}, Q_A]=0$ and, therefore, it is symmetric. 
The Rényi-$n$ entanglement asymmetry is then defined as the difference between the Rényi-$n$ entropies of $\rho_{A,Q}$ and $\rho_A$~\cite{amc-23},
\begin{equation}\label{eq:asymm}
    \Delta S^{(n)}_A = S^{(n)}(\rho_{A,Q})-S^{(n)}(\rho_A).
\end{equation}
The main properties of $\Delta S^{(n)}_A$ that make it a good quantifier of symmetry breaking in the subsystem $A$ are its positivity, $\Delta S^{(n)}_A \geq 0$, and the fact that $\Delta S^{(n)}_A = 0$ if and only if $[\rho_A,Q_A] = 0$. Asymmetry has been investigated in matrix product states~\cite{cv-24, mcp-25}, random states~\cite{ampc-24, ct-25, rac-24, rac-25}, and at~\cite{cc-23, fadc-24, frc-24, kmop-24, lmac-25, bgs-24} and away from~\cite{fac-24, cm-23, mff-25} critical points in many-body quantum systems, where it has been employed to study confinement, topological order, and stabilizerness~\cite{khor-24, ggh-25, tfhtp-24}. It has also appeared under different names in resource theories~\cite{vawj-08, gms-09} and quantum field theories~\cite{chmp-20, chmp-21}.

A particularly convenient way of writing the symmetrized reduced density matrix $\rho_{A, Q}$ is obtained by using the integral representation of the projector $\Pi_q$,
\begin{equation}
    \rho_{A,Q} = \int \frac{d\alpha}{2\pi} e^{i \alpha Q_A} \rho_A e^{-i\alpha Q_A}.
\end{equation}
Plugging this representation of $\rho_{A,Q}$ in Eq.~\eqref{eq:asymm}, it can be easily shown that the entanglement asymmetry can be rewritten as
\begin{equation}
\label{defasymmetry}
    \Delta S^{(n)}_A = \frac{1}{1-n} \log\left[\int_{[0,2\pi]^{n-1}} \frac{d\bm{\alpha}}{(2 \pi)^{n-1}} \frac{Z_n(\bm{\alpha})}{Z_n(\bm{0})}\right],
\end{equation}
where $\bm{\alpha} \equiv (\alpha_1,\cdots,\alpha_{n-1})$ and $Z_n(\bm{\alpha})$ are known as charged moments of $\rho_A$,
\begin{equation}
\label{defZn}
    Z_n(\bm{\alpha}) = \text{Tr}[\rho_A e^{i \alpha_1Q_A} \cdots\rho_Ae^{-i(\alpha_1+\cdots+\alpha_{n-1})Q_A}].
\end{equation}
In the following, we restrict ourselves to $n=2$ for practical reasons, although we expect the same qualitative behavior for any $n$, including in the limit $n\to 1$. In our case, the charge $Q_A$ of interest is the fermion number $Q_A=\sum_{j \in A} c^{\dagger}_jc_j$, which is quadratic in the fermionic operators. 

Our out-of-equilibrium protocol is the following: we prepare the system in an initial state $\ket{\psi_0}$ that breaks the $U(1)$ symmetry associated with particle number conservation; therefore, the entanglement asymmetry~\eqref{eq:asymm} is initially non-zero. We then evolve such a state with the $\nu=1$ QSSEP dynamics~\eqref{model_dH} and calculate the average asymmetry $\langle \Delta S_A(t)\rangle$.
We consider initial Gaussian states, since all the information about their time evolution is then fully contained in the two-point correlation functions $C_{nm}$ and $F_{nm}$, as we discussed in Sec.~\ref{sec:model}. Notice that, unlike in Sec.~\ref{sec:ent_fluct} where we considered initial symmetric states, here both $C_{nm}, F_{nm}\neq 0$ due to the breaking of particle number symmetry at $t=0$. In particular, we can choose as initial states  generic squeezed states, 
\begin{equation}\label{eq:squeezed_st}
\ket{\psi_0} \propto \exp \left(- \int \frac{dk}{2 \pi} \,\mathcal{M}_k c^{\dagger}_k c^{\dagger}_{-k}\right) \ket{0}, 
\end{equation}
with $\mathcal{M}_k$ an arbitrary real and odd function of $k$. 
As shown in Ref.~\cite{rylands-24}, the charged moments of~\eqref{eq:squeezed_st} are
\begin{equation}
\label{cmomsqu}
    Z_2(\alpha) = Z_2(0) \exp\left(-\ell\int \frac{dk}{2\pi} f_k(\alpha)\right),
\end{equation}
where the contribution per mode $f_k(\alpha)$,  
\begin{equation}
        f_k(\alpha) = - \log(1-n_k + n_k e^{2 i \alpha}),
\end{equation}
is determined by the density of occupied modes $n_k = \bra{\psi_0} c^{\dagger}_k c_k \ket{\psi_0} = \mathcal{M}_k^2/(1+\mathcal{M}_k^2)$. The entanglement asymmetry in these squeezed states follows directly from \eqref{cmomsqu} and \eqref{defasymmetry}. For large $\ell$, it behaves as
\begin{equation}
\label{DSt0}
    \Delta S^{(2)}_A(t = 0) = \frac{1}{2}\log\left(\pi \ell \int \frac{dk}{2\pi} n_k(1-n_k)\right).
\end{equation}

\begin{figure}[t]
\begin{subfigure}
    \centering
    \includegraphics[width=.5\linewidth]{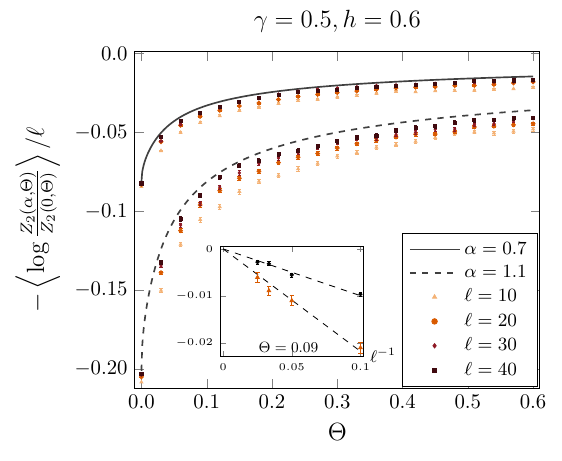}
\end{subfigure}
\begin{subfigure}
    \centering
    \includegraphics[width=.5\linewidth]{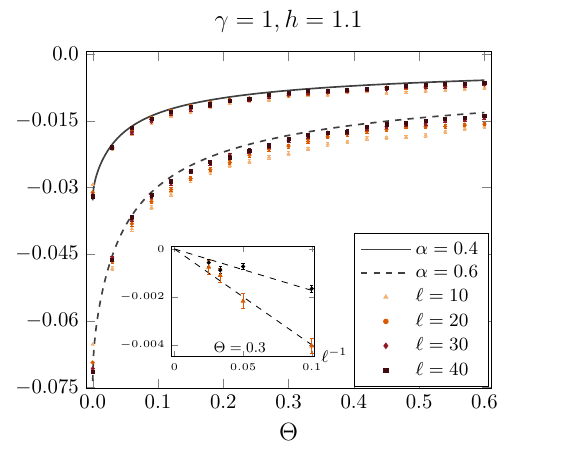}
\end{subfigure}
\caption{\label{fig:cmomfig} Left panel: Time evolution of  $-\langle\log Z_2(\alpha,t)/Z_2(0,t)\rangle$ after the quench from the ground state of~\eqref{xymodel} with $ (\gamma=0.5,h=0.6)$. The quench Hamiltonian is \eqref{model_dH}.  
This initial state breaks the particle-number symmetry. The solid and dashed curves correspond to the prediction of Eq.~\eqref{cmomdqp} for two different values of the parameter $\alpha$. The symbols are the average of the exact charged moment $Z_2(\alpha,t)$ computed numerically over approximately $10^3$ noise realizations, for different subsystem sizes $\ell$ in the thermodynamic limit $L \to \infty
$. The errorbar of each symbol is the standard deviation of the mean of the set of collected samples. Right panel: Same as the left but for initial Hamiltonian with with $\gamma = 1 $ and $h=1.1 $, and   $\alpha=0.4,0.6$. In the insets, we subtract, at a fixed $\Theta$, the prediction \eqref{cmomdqp} from the exact numerical data shown in the main plots.  The result is plotted as a function of  $1/\ell$. The linear fits (dashed lines) show that the leading correction  is  $\sim \ell^{-1}$.}
\end{figure}

Let us first study the averaged time evolution of the charged moments. It has been shown that, in a quench from~\eqref{eq:squeezed_st} in the non-random case, $\log Z_2(\alpha, t)$ admits a quasiparticle expression of the form of Eq.~\eqref{eq:z_F}. We can then apply for it the same recipe as for the R\'enyi entanglement entropy in Sec.~\ref{sec:ent_fluct}. Precisely, the quasiparticle prediction for the dynamics in the $\nu=1$ QSSEP is obtained by replacing the distance between the quasiparticles in a pair by the random variable $2|\xi_k|$. We then have the formal expression
\begin{equation}\label{eq:cm_qp}
-\log \frac{Z_2(\alpha,t)}{Z_2(0,t)} \asymp \ell \int \frac{dk}{2\pi} \,f_k(\alpha)\, x_{\xi_k}. 
\end{equation}
By taking   the noise average $\langle\cdot\rangle$ and using the same logic as for the R\'enyi entanglement entropy, we find
\begin{equation}
\label{cmomdqp}
    - \left\langle \log\frac{Z_2(\alpha,t)}{Z_2(0,t)}\right\rangle = \ell \left(\int \frac{dk}{2\pi} f_{\alpha}(k)\right) \int d\xi \,P(\xi,t)\, \text{max}(1-2|\xi|/\ell,0).
\end{equation}
 We observe that,  as for the noise-averaged entanglement entropy~\eqref{avgS}, pairs with different momenta are uncorrelated in Eq.~\eqref{cmomdqp}, and the dynamics does not depend on the momentum. The effect of noise-induced correlations between  pairs  with different momenta will instead play a role in the dynamics of the entanglement asymmetry. In Fig.~\ref{fig:cmomfig}, we compare the prediction~\eqref{cmomdqp} with exact numerical results. The numerical data converge to Eq.~\eqref{cmomdqp} as the subsystem length $\ell$ increases, thereby confirming the validity of the diffusive quasiparticle picture for the noise-averaged charged moments. 
 In the figure, we take as symmetry-breaking initial states the ground state of the Kitaev chain,
\begin{equation}
\label{xymodel}
    H(\gamma,h) =-\frac{1}{2} \sum_j (c^{\dagger}_j c_{j+1} + \gamma c^{\dagger}_j c^{\dagger}_{j+1} + 2h c^{\dagger}_jc_j+h.c.), 
\end{equation}
for different values of the parameters $\gamma$ and $h$. For $\gamma\neq 0$, this Hamiltonian does not commute with the total particle number operator $Q$ and, therefore, the ground state breaks the corresponding $U(1)$ symmetry and is of the form~\eqref{eq:squeezed_st}. The evolution of the charge moments 
and the entanglement asymmetry after a quench from these initial states in the non-random case has been studied in Ref.~\cite{makc-24} using the quasiparticle picture.  The specific form of the function $f_k(\alpha)$ for these initial states reads~\cite{makc-24} 
\begin{equation}
    f_k(\alpha) = -\frac{1}{2}\log(1- \sin^2 \alpha \sin^2 \Delta_k),
\end{equation}
with 
\begin{equation}
    \sin \Delta_k = \frac{\gamma \sin k}{\sqrt{(h-\cos k)^2 + \gamma^2 \sin^2 k}}.
\end{equation}

 We now analyze the time evolution of the entanglement asymmetry. First, by combining the conjecture in Eq.~\eqref{generaldqp} with Eqs.~\eqref{defasymmetry} and~\eqref{eq:cm_qp}, we expect that the average Rényi-2 asymmetry is described in the diffusive scaling limit by
\begin{equation}
\label{DSdiffQP}
    \langle \Delta S^{(2)}_A(\Theta)  \rangle =-  \left\langle\, \log\left[ \int \frac{d\alpha}{2 \pi} \exp\left(-\ell \int \frac{dk}{2 \pi} x_{\xi_k}f_k(\alpha)\right)\right] \right\rangle_{\Theta}.
\end{equation}
Here the average is over different realizations of the random variable $W_t = W_{R,t} + i W_{I,t}$, with $W_{R(I),t}$ normally distributed variables with distribution $P(\xi,t)$, from which one constructs $\xi_k$  and $x_{\xi_k}$ (cf.~\eqref{eq:xi-def} and~\eqref{eq:x-def}).
According to Eq.~\eqref{DSdiffQP}, only the quasiparticle pairs that are fully contained in the subsystem $A$, see Fig.~\ref{fig:DQPsketch}, contribute to $\langle \Delta S_A^{(2)}(\Theta)\rangle$. Since the probability distribution $P(\xi, t)$ that governs the separation $\xi_k(t)$ between the two quasiparticles of a pair becomes wider with $t$, the probability of having entangled pairs fully contained in $A$ decreases with time, vanishing in the long time limit. As a consequence, Eq.~\eqref{DSdiffQP} predicts that $\langle \Delta S_A^{(2)}(\Theta)\rangle$ decreases in time and eventually tends to zero when $\Theta\to\infty$. This means that, in the stationary state, the particle number symmetry, which was initially broken, is on average restored within subsystem $A$. 

\begin{figure}[t]
\begin{subfigure}
    \centering
    \includegraphics[width=.5\linewidth]{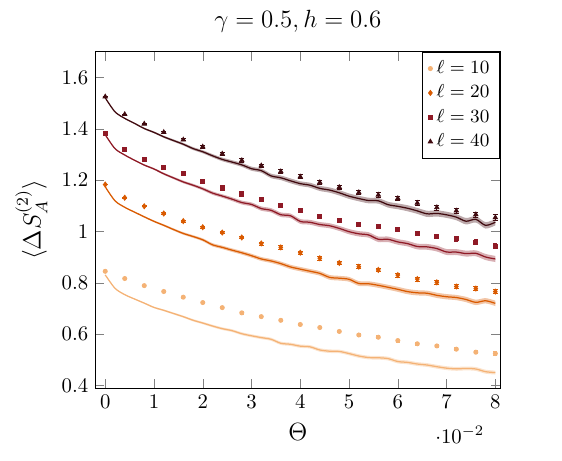}
\end{subfigure}
\begin{subfigure}
    \centering
    \includegraphics[width=.5\linewidth]{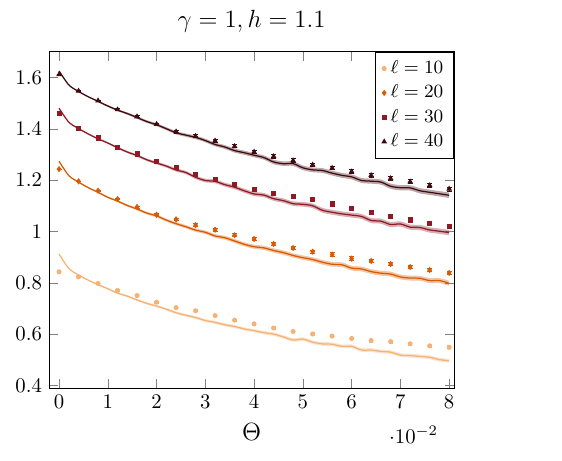}
\end{subfigure}
\caption{\label{fig:earlytimeDS} Left panel: Early time evolution of the average Rényi-2 entanglement asymmetry after a quench in the $\nu = 1$ QSSEP model from the ground state of Hamiltonian~\eqref{xymodel} with $h=0.5$ and $\gamma=0.6$. The symbols are the result of averaging the exact entanglement asymmetry over  $10^3$ different noise realizations for increasing subsystem sizes $\ell$ in the thermodynamic limit $L\to\infty$. The solid curves show the prediction of the diffusive quasiparticle picture in Eq.~\eqref{DSdiffQP}, where the average has been estimated by numerical sampling, and the shaded region represents the confidence interval of the mean value $\langle \Delta S^{(2)}_A\rangle \pm 3 \sigma$.  Right panel: Same as the left but considering as initial state the ground state of Eq.~\eqref{xymodel} with $\gamma = 1 $ and $h=1.1 $.}
\end{figure}

In Fig.~\ref{fig:earlytimeDS}, we compare the prediction of the quasiparticle picture in Eq.~\eqref{DSdiffQP} with the exact numerical data for the R\'enyi-2 asymmetry. The data are obtained by sampling over different noise realizations, as explained in Appendix~\ref{appA}. The initial states are the same used to discuss the charged moments in Fig.~\ref{fig:cmomfig}; i.e.  the ground state of the Kitaev chain~\eqref{xymodel} for two different values of $\gamma$ and $h$. In the plots, we restrict ourselves to ``small'' $\Theta\approx 0.1$, where the decay of the average asymmetry to zero is not yet apparent. Indeed, due to the diffusive scaling  of the asymmetry, to observe the symmetry restoration we would need to consider $t\gg \ell^2$. For this reason, beyond the times shown in the plot, the exact numerical calculation of the time-evolved correlation matrices becomes increasingly challenging. 
Nevertheless, we can see that, as the subsystem size $\ell$ increases, the numerical data converge to the values predicted by Eq.~\eqref{DSdiffQP}. 

Although Eq.~\eqref{DSdiffQP} is hard to treat analytically due to the presence of the logarithm inside the expectation value, we can extract from it the analytical long-time behavior of the asymmetry and its dependence on the initial state. This analysis is crucial to understand the time scales of the symmetry restoration and the presence of the quantum Mpemba effect.

\begin{figure*}[t]
\centering
\includegraphics[width=0.6\textwidth]{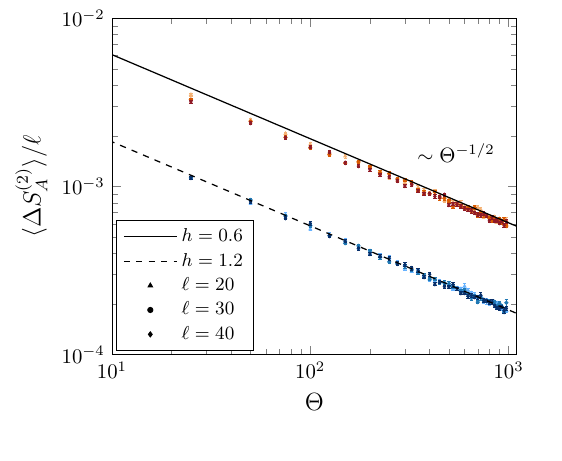} 
\caption{\label{fig:asymlatetime} Check of the result in Eq.~\eqref{largetimeDS} for the long time behavior of the Rényi-2 entanglement asymmetry for two different symmetry-breaking initial states: the ground state of the Hamiltonian~\eqref{xymodel} with $h = 0.6$ (solid line) and $h  =1.2$ (dashed line) and the same $\gamma = 0.5$. The symbols correspond to Eq.~\eqref{DSdiffQP}, where the average has been estimated numerically,
for several subsystem sizes $\ell$. We plot in the {\it y}-axis the ratio $\langle \Delta S^{(2)}\rangle/\ell$, observing a collapse of the symbols for the different $\ell$, as predicted by Eq.~\eqref{largetimeDS}.}
\end{figure*}

For convenience, we write the complex Brownian motion $W_t$ that enters in $\xi_k(t)$ in polar coordinates, i.e. as $W_t = \rho e^{i\varphi}$. The probability distribution of $\rho$ at time $t$ is $P_t(\rho) = e^{-\rho^2/(Dt)} 2\rho/(Dt)$ while the phase $\varphi$ is at each time distributed uniformly over $[-\pi,\pi)$ independently on the stochastic variable $\rho$. The counting factor $x_{\xi_k}$ can then be rewritten as $x_{\xi_k} = \text{max}(1-4\rho |\sin(k-k^*)|/\ell,0)$, where $k^* = \pi/2-\varphi$ is also uniformily distributed. At long times, $\Theta \gg 1$, realizations with $ x_{\xi_k} \neq 0 $ are statistically suppressed due to the time evolution of the probability distribution $P_{t}(\rho)$. For this reason, in that regime, we can expand the exponential and, subsequently, the logarithm in Eq.~\eqref{DSdiffQP}, obtaining
\begin{equation}
\label{eq:large_t_asymm}
\langle \Delta S^{(2)}_A(\Theta)  \rangle \underset{\Theta\gg 1}{\simeq} \ell\,\left\langle \int\frac{dk}{2\pi} g_k  \,\text{max}\left(1-4\rho |\sin(k-k^*)|/\ell,0\right)\right\rangle_{\Theta},
\end{equation}
where 
\begin{equation}
g_k \equiv \int \frac{d\alpha}{2\pi} f_k(\alpha).
\end{equation}
The asymptotic vanishing behavior of  Eq.~\eqref{eq:large_t_asymm} is determined by the modes close to $k^*$ and $k^* + \pi$, i.e. the ones that satisfy $|\sin(k-k^*)|<\ell/(4 \rho)$. Since at large times $\rho/\ell \sim \sqrt \Theta$ (on average), the sine in Eq.~\eqref{eq:large_t_asymm} can be expanded around $k\simeq k^*$ and $k\simeq k^*+\pi$, and we obtain
\begin{equation}
\label{expDS}
    \langle \Delta S^{(2)}_A(\Theta)  \rangle \underset{\Theta\gg 1}{\simeq} 2 \ell \left\langle
    \int_{k^*-\ell/(4 \rho)}^{k^*+\ell/(4 \rho)}\frac{dk}{2\pi}g_k \left(1-\frac{4 \rho}{\ell} |k-k^*|\right) 
\right\rangle_{\Theta},
\end{equation}
where the factor $2$ takes into account the contribution of both $k \simeq k^*$ and $k\simeq k^*+\pi$, which are equal.
It is important to remark that, to obtain~\eqref{eq:large_t_asymm} we performed an expansion in Eq.~\eqref{DSdiffQP} in which we neglected the contribution from realizations with small values of $\rho$. Therefore, to make the result in Eq.~\eqref{expDS} self-consistent when computing the expectation value, we need to introduce a cutoff $\epsilon$ on the smallest allowed value of $\rho$.
For this reason, we compute the expectation value~\eqref{expDS} as follows
\begin{equation}
    \langle \Delta S^{(2)}_A(\Theta)  \rangle \underset{\Theta\gg 1}{\simeq} 2 \ell  \int_{\epsilon}^{\infty} d\rho \,P_t(\rho)\left \langle \int_{k^*-\ell/(4 \rho)}^{k^*+\ell/(4 \rho)} \frac{dk}{2\pi} g_k \, \left(1- \frac{4 \rho}{\ell} |k-k^*|\right) \right \rangle_{k^*},
\end{equation}
where $\langle \cdot \rangle_{k^*}$ denotes the uniform average over $k^*$. Expanding $g_k$ up to second order around $k^*$ and evaluating the integrals in $\rho$ and $k$ yields, after some straightforward algebra,
\begin{equation}
    \langle \Delta S^{(2)}_A(\Theta)  \rangle \underset{\Theta\gg 1}{\simeq}  \frac{\ell \, \langle g_{k^*} \rangle}{4\sqrt{\pi}} \frac{1}{\sqrt{\Theta}} \text{erfc} \left (\frac{\epsilon}{\sqrt{D t}}\right) + O\left(1/\epsilon,1/\Theta \right).
\end{equation}
In this expression, we first take the limit of large~$\Theta$, thus neglecting the subleading terms, and then send the cutoff~$\epsilon$ to zero, obtaining the finite result
\begin{equation}
\label{largetimeDS}
     \langle \Delta S^{(2)}_A(\Theta)  \rangle \underset{\Theta\gg 1}{\simeq} \frac{\ell \,\Upsilon} {\sqrt{\Theta}},
\end{equation}
where the proportionality constant is $\Upsilon =  \langle g_{k^*}\rangle/(4 \sqrt{\pi})$, and the average of the uniformly distributed variable $k^*$ is over $[-\pi,\pi)$. Therefore, $\Upsilon$ only depends on the 
density of occupied modes $n_k$ of the initial state. In the case of an initial squeezed state~\eqref{eq:squeezed_st}, the function $g_k$ is given by \cite{rylands-24}
\begin{equation}
\label{funcgk}
    g_k = - \text{max}(\log n_k, \log(1-n_k)).
\end{equation}
In Fig.~\ref{fig:asymlatetime}, we check the prediction in Eq.~\eqref{largetimeDS} for the long time behavior of the average Rényi-2 asymmetry against Eq.~\eqref{DSdiffQP}. In that figure, we plot $\langle\Delta S_A^{(2)}\rangle/\ell$ versus $\Theta$ for several values of $\ell$. We consider the dynamics from the ground state of the model \eqref{xymodel} with $\gamma = 0.5$ and $h=0.6$ or $h=1.2$. As it is clear from Fig.~\ref{fig:asymlatetime}, the data for different $\ell$ collapse on the same curve, as expected because the asymmetry is function of $\Theta$ and of $t$ and $\ell$ separately. Moreover, $\langle \Delta S_A^{(2)} \rangle/\ell$ is $\sim \Theta ^{-1/2}$ in the limit $\Theta\to\infty$. 
The continuous and the dashed-dotted lines in Fig.~\ref{fig:asymlatetime} are Eq.~\eqref{largetimeDS}

We can now compare our result with the known behavior of the entanglement asymmetry for the same initial states and with non-random hopping amplitudes, studied in Ref.~\cite{makc-24}. In both cases, the (average) asymmetry decays to zero in time, indicating the dynamical restoration of the symmetry in the stationary state. However, in the $\nu$-QSSEP model the decay is $\sim 1/\sqrt{t}$ due to the diffusive nature of the dynamics, whereas in the non-random case it is $1/t$ or $1/t^3$, since the quasiparticles propagate ballistically. Moreover, the prefactor $\Upsilon$ (cf.~\eqref{largetimeDS}) that encodes the information of the initial state is given by $g_k$ (cf.~\eqref{funcgk}) in both cases, but with a key difference. In the non-random case, this prefactor is, up to a proportionality constant, the value of the function $g_k$ at the slowest mode. In the $\nu$-QSSEP, instead, the prefactor is the average of $g_k$ equally weighted over all possible momenta. That is, the restoration of symmetry at long times is governed by the slowest mode in the non-random case, whereas all modes contribute when the hopping amplitudes are random.
This difference is crucial for the existence of the quantum Mpemba effect. 

Indeed, the quantum Mpemba effect occurs when, for a pair of initial states, the symmetry is restored faster in the dynamics from the state that initially breaks it more.
Let us consider, therefore, two initial states of the form in Eq.~\eqref{eq:squeezed_st}, denoted $1$ and $2$, that break the particle number symmetry. Therefore,
let us assume that at $t = 0$, $\Delta S_{A,1}(0)>\Delta S_{A,2}(0)$, where $\Delta S_{A,1(2)}(0)$ is the initial asymmetry in the state $1(2)$. Using the result in Eq.~\eqref{DSt0}, this condition can be rewritten in terms of the occupation functions of the two initial states $n^{(1,2)}_k$ as
\begin{equation}
\label{condition1}
    i)\quad \int \frac{dk}{2 \pi} n_k^{(1)}(1-n^{(1)}_k) > \int \frac{dk}{2 \pi} n_k^{(2)}(1-n^{(2)}_k).
\end{equation}
The quantity $n_k(1-n_k)/(2\pi)$ is also known as the charge susceptibility per mode~\cite{rylands-24}. The quantum Mpemba effect occurs when
there is a time after which  $\Delta S_{A,1}(t)<\Delta S_{A,2}(t)$ for all $t$.  
According to Eq.~\eqref{largetimeDS}, which describes the long time behavior of the average asymmetry, this occurs when $\Upsilon_1 < \Upsilon_2$. Using the explicit expression of $\Upsilon$ in Eq.~\eqref{largetimeDS}, we can also rewrite this condition in terms of the occupation functions $n^{(1,2)}_k$ as follows
\begin{equation}
\label{condition2}
    ii)\quad - \int \frac{dk}{2 \pi} \text{max}(\log n^{(1)}_k, \log(1-n_k^{(1)})) < - \int \frac{dk}{2 \pi} \text{max}(\log n^{(2)}_k, \log(1-n^{(2)}_k)).
\end{equation}
Given two initial squeezed states with different asymmetry, Eqs.~\eqref{condition1} and~\eqref{condition2} are the necessary and sufficient conditions for the existence of quantum Mpemba effect in the $\nu=1$ QSSEP. Notice that they only depend on the density of occupied modes of the initial states.

We numerically investigated the conditions~\eqref{condition1} and~\eqref{condition2}  for several initial states. We observed that the presence of the  momentum average in~\eqref{condition2}, in sharp contrast to the non-random case in which the late-time behavior is determined by the slowest mode, inhibits the Quantum Mpemba effect. In fact, we studied the conditions~\eqref{condition1} and~\eqref{condition2} for pairs of ground states of~\eqref{xymodel} corresponding to different $h$ and $\gamma$ and for which the Mpemba effect occurs in the non-random case, finding that most of them violate condition~(ii).

\begin{figure*}[t]
\centering
\includegraphics[width=0.6\textwidth]{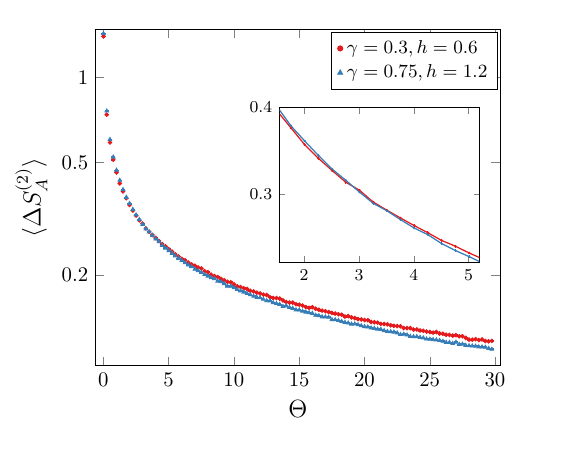} 
\caption{\label{fig:crossing} Example of the occurrence of the quantum Mpemba effect as predicted by the conditions~\eqref{condition1} and~\eqref{condition2} for two different initial states corresponding to the ground state of the Hamiltonian~\eqref{xymodel} for $(\gamma = 0.3, h = 0.6)$ (red circles) and for $(\gamma = 0.75, h = 1.2)$ (blue triangles). The symbols are obtained numerically from the quasiparticle prediction in Eq.~\eqref{DSdiffQP} for a subsystem of length $\ell = 60$.
The inset shows a zoom of the time interval where the entanglement asymmetries cross.
Both plots are in log scale on the vertical axis. 
In the $\nu=1$ QSSEP, the appearance of this phenomenon is restricted to the case of nearly close initial asymmetries.}  
\end{figure*}

Precisely, our analysis suggests that Mpemba effect occurs only for initial states with nearly identical asymmetries. In Fig.~\ref{fig:crossing} we provide such an example in which the Mpemba effect occurs. The values of the asymmetry for short and long times undoubtedly show the presence of the Mpemba effect, but the determination of the exact location of the crossing point is clearly very difficult even numerically. The figure evidences that the quantum Mpemba effect in the $\nu=1$ QSSEP is a highly fine tuned phenomenon.
For instance, the Mpemba effect does not occur for the tilted ferromagnetic states $e^{-i\theta/2\sum_j\sigma_j^y}\ket{\uparrow\cdots\uparrow}$, which are the ground state of the Hamiltonian~\eqref{xymodel} along the curve $h^2+\gamma^2=1$ with $\cos^2\theta=(1-\gamma)/(1+\gamma)$. This is remarkable, as the quantum Mpemba effect is typically observed for any pair of initial states within this family — not only in the non-random analogue of the $\nu = 1$ QSSEP~\cite{amc-23}, but also in random unitary circuits~\cite{tcd-24} and in the experiment~\cite{joshi-24}. The asymmetry of these states is $\Delta S_A^{(2)}=1/2\log(\ell/2\sin^2\theta)$ and, therefore, it increases monotonically with $\theta\in[0,\pi/2]$. On the other hand, we numerically verified that, for these states, $\Upsilon(\theta_1)<\Upsilon(\theta_2)$ for any $0\leq \theta_1<\theta_2\leq\pi/2$ and, therefore, the condition (ii) for the occurence of the Mpemba effect is never satified.

\section{Conclusions}\label{sec:conclusions}

In this work, we studied the out-of-equilibrium dynamics of the statistics of entanglement in a spatially homogeneous version of the Quantum Symmetric Simple Exclusion Process (QSSEP) model, the $\nu=1$ QSSEP. We built upon the results of Ref.~\cite{alba-25}, in which it is  shown that the average entanglement growth is diffusive, and it can be described by the propagation of entangled pairs of quasiparticles that perform a Brownian motion. We extended this picture by taking into account the statistical correlations between the different pairs of quasiparticles. This generalization allows us to exactly characterize the full-time dynamics of all the higher moments and, therefore, the full distribution probability of the entanglement entropy.

The diffusive quasiparticle picture developed here can be applied to calculate the average full-time dynamics of any sufficiently well-behaved function of the reduced density matrix of a subsystem in the $\nu=1$ QSSEP model. In particular, we used it to analyze the dynamical restoration of the symmetry associated with particle number conservation, and the emergence of the quantum Mpemba effect. To this end, we derived the exact time evolution in the diffusive scaling limit of the average entanglement asymmetry, which measures the degree of symmetry breaking in a subsystem. In contrast to the analogous non-random model, we showed that the stochasticity of the time evolution strongly inhibits  the quantum Mpemba effect. This result is in contrast with what has been observed in random unitary circuits~\cite{tcd-24,lzyz-24}. 

There are several interesting directions for future work. 
For instance, it would be interesting to further investigate the role of stochastic dynamics in the Mpemba effect.
An immediate question is whether the dynamics of the generic $\nu-$QSSEP model, i.e. when spatial invariance is only partially recovered, admits a similar diffusive quasiparticle picture. While in the fully inhomogeneous limit $\nu \to \infty$, the entanglement generation mechanism is likely related to the one discussed in Ref.~\cite{sbn-25}, for a finite value $\nu\geq2$, the mechanism is not understood yet. For this reason, studying the model at finite $\nu$ could provide further insight into the crossover between those two extreme cases.
Another possibility consists of generalizing the $\nu-$QSSEP model to include interactions. In generic interacting models, the linear growth of entanglement is expected to be restored \cite{knap-18,rl-18, kh-13}; however, a stochastic continuous-time Hamiltonian dynamics in which the generator is an interacting integrable model could instead retain a diffusive entanglement spreading. Developing a diffusive analogue of the quasiparticle picture for interacting integrable systems~\cite{ac-17} is  an interesting direction. 
\newline

\textbf{Acknowledgments.}
AR and FA thank Shion Yamashika, Konstantinos Chalas and Colin Rylands for discussions. 
PC, FA and AR acknowledge support by the ERC-AdG grant MOSE No. 101199196. This work has been supported by the project “Artificially devised many-body quantum dynamics in low dimensions - ManyQLowD” funded by the MIUR Progetti di Ricerca di Rilevante Interesse Nazionale (PRIN) Bando 2022 - grant 2022R35ZBF.

\appendix
\section{Numerical techniques}
\label{appA}

In this Appendix, we outline the methods used to compute the 
time evolution of the correlation matrices under Eq.~\eqref{eq:time_ev} for 
different noise realizations, which serve to validate the results 
presented in the main text. Let us discuss separately the case of initial states that have particle-number symmetry but break translational invariance, studied in Sec.~\ref{sec:ent_fluct}, from those that are translationally invariant but break particle-number symmetry, considered in Sec.~\ref{sec:asymmetry}.

\textbf{Quench dynamics from the Néel state}: In this case, $F_{nm}(t)=0$ for all $t$. Taking the correlation matrix at $t=0$, Eqs.~\eqref{eq:time_ev} and~\eqref{eq:corr_mat}, and evolving it with $dH_t$ in the Fourier representation~\eqref{dhmomspace}, we have that at time $dt$, 
\begin{equation}\label{eq:G_dt}
G_{nm}(dt)=\int \frac{dk}{2 \pi} e^{ik(n-m)}\left[\frac{f(k)}{2}+e^{i\pi m}e^{i2d\varepsilon_{0, k}}\frac{g(k)}{2}\right],
\end{equation}
where $d\varepsilon_{0, k}=2 \sqrt{D} \,\text{Re}\{dW_0 e^{i k}\}$. Iterating Eq.~\eqref{eq:G_dt}, and applying the fact that $[dH_t, dH_{t’}]=0$ for any $t\neq t’$, the correlation matrix $G(t)$ at time $t$ is given by
\begin{equation}\label{eq:num_G}
G_{nm}(t)=\int \frac{dk}{2 \pi} e^{ik(n-m)}\left[\frac{f(k)}{2}+e^{i\pi m}e^{i2\varepsilon_{k}(t)}\frac{g(k)}{2}\right],
\end{equation} 
where $\varepsilon(t)=2 \sqrt{D}\, \text{Re}\{W_t\,e^{ik}\}$ and $W_t$ is the sum of the increments $dW_{t’}$ with $t’\in [0, t)$.  For the numerical calculations, we generate the random hoppings
$dW_t = dW_{R, t} + idW_{I,t}$, where $dW_{R, t}$ and $dW_{I, t}$ are real and normally distributed with
zero mean and variance $Ddt/2$, taking  $dt = 0.05$. We verified that the result of the computation does not significantly change when decreasing $dt$. From the restriction $G_A$ of the correlation matrix~\eqref{eq:num_G} to $A$, the entanglement entropy can be calculated using the formula~\cite{pe-09}
\begin{equation}
    S_A = -\text{Tr}\left[ G_{A} \log G_A + (\mathbb{I}-
    G_A) \log (\mathbb{I}-
    G_A) \right].
\end{equation}

\textbf{Quench dynamics from a symmetry breaking initial state}:
In this case, we can apply the same strategy as before, but now the correlations $F_{nm}$ are not zero in the initial state. Therefore, it is more natural to arrange all the two-point correlations in the single matrix
\begin{equation}
    \Gamma_{nm} = 2 \left(\begin{array}{cc} G_{nm} & \overline{F_{mn}} \\ F_{nm} & \delta_{nm}-G_{mn}
\end{array}\right) - \delta_{nm}.
\end{equation}
Since both the initial state and the time evolution are translationally invariant, we can write $\Gamma$ as
\begin{equation}
    \Gamma_{nm}(t) = \int \frac{dk}{2 \pi}  e^{i k(n-m)} \mathcal{C}(k,t),
\end{equation}
where $C(k,t)$ is a $2\times 2$ matrix. In particular, when the initial state is the ground state of the Kitaev chain~\eqref{xymodel}, it reads
\begin{equation}
    \mathcal{C}(k,t) = \begin{pmatrix}
        \cos \Delta_k &i e^{-2 i \varepsilon_k(t)} \sin \Delta_k \\
        -i e^{2 i \varepsilon_k(t)} \sin \Delta_k & -\cos \Delta_k,
    \end{pmatrix},
\end{equation}
with 
\begin{equation}
    \cos \Delta_k = \frac{h - \cos k}{\sqrt{(h - \cos k)^2+\gamma^2 \sin^2 k}},
\end{equation}
\begin{equation}
    \sin \Delta_k = \frac{\gamma \sin k}{\sqrt{(h - \cos k)^2 + \gamma^2 \sin^2 k}}.
\end{equation}
The random variable $\varepsilon_k(t)$ is generated as in the case of symmetric states, and we take again $dt=0.05$. The charged moments $Z_n(\boldsymbol{\alpha})$, defined in Eq.~\eqref{defZn}, can be calculated from the restriction $\Gamma_A$ of $\Gamma$ to $A$ with the formula~\cite{amc-23} 
\begin{equation}
    Z_n(\bm{\alpha}) = \left( \text{det}\left[ \left(\frac{\mathbb{I}-\Gamma_A}{2}\right)^n \left( \mathbb{I}+ \prod_{j=1}^n W_j\right) \right]\right)^{1/2},
\end{equation}
where $W_j = (\mathbb{I}+\Gamma_A)(\mathbb{I}-\Gamma_A)^{-1} e^{i \alpha_j n_A}$ and $\alpha_n = -\sum_{j=1}^{n-1}\alpha_j$. The matrix $n_A$ is a diagonal $2\ell\times 2\ell$ matrix with entries $(n_A)_{2j-1, 2j-1}=-1$ and $(n_A)_{2j, 2j}=1$, $j=1, \dots, \ell$.
Once the charged moments are determined numerically, the entanglement asymmetry can be calculated applying Eq.~\eqref{defasymmetry}.

\end{document}